# Coupled multiphysics analysis of a 4-vane RFQ accelerator under high power operation

Xiaowen Zhu*, Claude Marchand, Olivier Piquet, Michel Desmons

*IRFU, CEA, Université Paris-Saclay, F-91191 Gif-sur-Yvette, France*

**Abstract:**

The radiofrequency resonant four-quadrant mechanical structure of a 4-vane Radio Frequency Quadrupole (RFQ) has a high quality factor and narrow bandwidth, resulting in high sensitivity to frequency detuning caused by thermal expansion under high power operation, so minimization of frequency errors and affording dynamic tuning are very important design issues. Here we describe an optimization approach to figure out a suitable cooling design for a 4-vane RFQ in steady state. Besides, we investigate how the accelerator responds in transient thermal analysis that could help to guide commissioning and reduce frequency detuning. Multiphysics analysis utilized with CST and ANSYS for a recently developed 176 MHz 4-vane RFQ is taken as an example. This RFQ will dissipate 211 kW when reaching an inter-vane voltage of 80 kV, which is required for an acceleration of an 80-mA proton from 65 keV to 2.5 MeV in 5.3 meters.

## 1. Introduction

The Radiofrequency Quadrupole (RFQ) plays a vital role in almost all ion linear accelerators, including front-end injectors for acceleration of high intensity low energy ion beam in both high-energy accelerator facilities [1,2,3,4,5] and low-energy compact accelerator-based neutron sources (CANS) [6,7,8,9,10]. It effectively fulfils the demands of beam bunching and acceleration in the longitudinal phase space, while providing alternating focusing forces to a beam in transverse phase spaces. Especially for high power operation, a 4-vane RFQ is very favorable, presenting advantages in high mechanical strength, easy heat removal and low dipole field component, thus preserving high transmission efficiency and high beam quality to downstream accelerators. However, a 4-vane RFQ is a resonating structure with high quality factor, which gives a small bandwidth and high sensitivity during high power operation. The RF power dissipation on the structure will heat the wall and deform the cavity, which leads to a frequency detuning in the RFQ. For proper acceleration of a beam in the downstream accelerator sections, the RFQ operating frequency must be tuned to the designed one. The common solution is to mitigate the frequency detuning during high power operation by adjusting the coolant temperature inside the RFQ accelerator [11,12,13,14,15,16,17] through a resonance control cooling system (RCCS).

In this paper, we propose a way to optimize the layout of cooling channels of a 4-vane RFQ, where the inputs of tuning coefficients of vane and wall channels required for frequency regulation are determined. In addition, the transient frequency response during start-up when the heat load applied to RFQ cavity changes significantly has been evaluated to decrease frequency detuning. In the coming sections, we will take a high power 176 MHz 4-vane RFQ design for a new high brilliance neutron source project at CEA Paris-Saclay as an example [18]. The specifications of this design are shown in Table 1. This RFQ is designed to accelerate an 80-mA proton beam from 65 keV to 2.5 MeV with a duty factor of 10%. The 5.3-meter long RFQ is a 4-vane structure and is divided into 5 sections. The nominal input power is 211 kW for a tip-to-tip voltage of 80 kV.

*Corresponding author at: IRFU, CEA, Université Paris-Saclay, F-91191 Gif-sur-Yvette, France.
*E-mail address:* zhuxw13@pku.edu.cn (X.W. Zhu)

Table 1 Main design parameters of RFQ

| Parameters | Proton RFQ |
|---|---|
| Ion species | $H^+$ |
| Resonant frequency (MHz) | 176 |
| Peak beam current (mA) | 80 |
| Injection energy (keV) | 65 |
| Extraction energy (MeV) | 2.5 |
| Inter-vane voltage (kV) | 80 |
| Accelerator length (cm) | 529.0 |
| Copper power (kW, plus 20% margin) | 211 |
| Kilpatrick limit | 1.403 |
| Transmission efficiency (%) | 98.9 |

## 2. Multiphysics Procedure

The multiphysics analysis is a coupled electromagnetic, thermal and mechanical analysis based on Finite Element Method (FEM). Both CST [19] and ANSYS [20] code packages are implemented in the multiphysics simulations. Figure 1 shows the flow chart for coupled-field analysis of a high frequency structure. The surface power loss of inner cavity wall is obtained from electromagnetic simulation with electromagnetic boundary conditions. Next, this power loss distribution is sent to the multiphysics model, but the power level must be scaled to the designed one. Then, associated thermal boundary conditions and heat convection coefficients are applied to evaluate the temperature map. After that, the acquired temperature distribution is imported by structural analysis with proper mechanical constrains for the field map calculations of yield stress and deformed mesh. Finally, the deformed mesh is returned back to the electromagnetic analysis to obtain the corresponding frequency detuning.

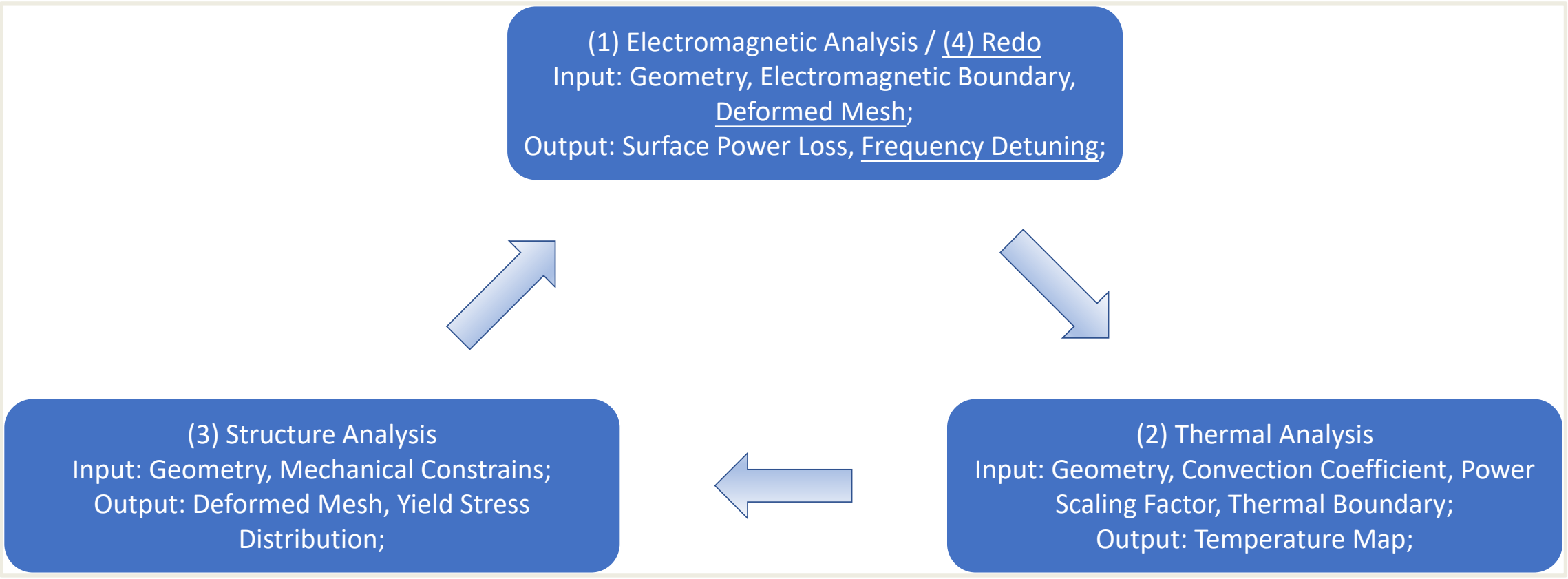

Fig.1 Scheme of multiphysics analysis used in CST and ANSYS codes.

To shorten simulation time and reduce memory usage, only a quarter of the 3D slice (10 mm) RFQ model of the final cross-sectional profile is put into simulations and the heat convection coefficients of cooling channels are calculated by empirical formulas [21]. For a circular pipe, the heat convection coefficient $h$ can be expressed as,

$$h = \frac{N_u K}{d} \qquad (1)$$

where $K$ represents the thermal conductivity of coolant, $d$ is the inner diameter of the cooling pipe, and $N_u$ is the Nusselt number. For the case of heating coolant, it can be given by,

$$N_u = 0.023 R_e^{0.8} P_r^{0.4} \quad (2)$$

The dimensionless Reynolds number is denoted by,

$$R_e = \frac{\rho \bar{v} d}{\mu} \quad (3)$$

where $\rho$ is the coolant density, $\bar{v}$ is the average flow rate, and $\mu$ is the coolant viscosity. A high Reynolds number (> 4000) would lead to a turbulent flow in a cooling pipe.

Moreover, the Prandtl number also is a dimensionless number, which measures the efficiency of transport by momentum diffusivity to thermal diffusion of a liquid and is written as,

$$P_r = \frac{C_p \mu}{K} \quad (4)$$

where $C_p$ is the specific heat capacity. The thermophysical characteristics of water as a function of temperature at atmospheric pressure could be found in the Ref. [22].

## 3. Optimization of Channel Configuration

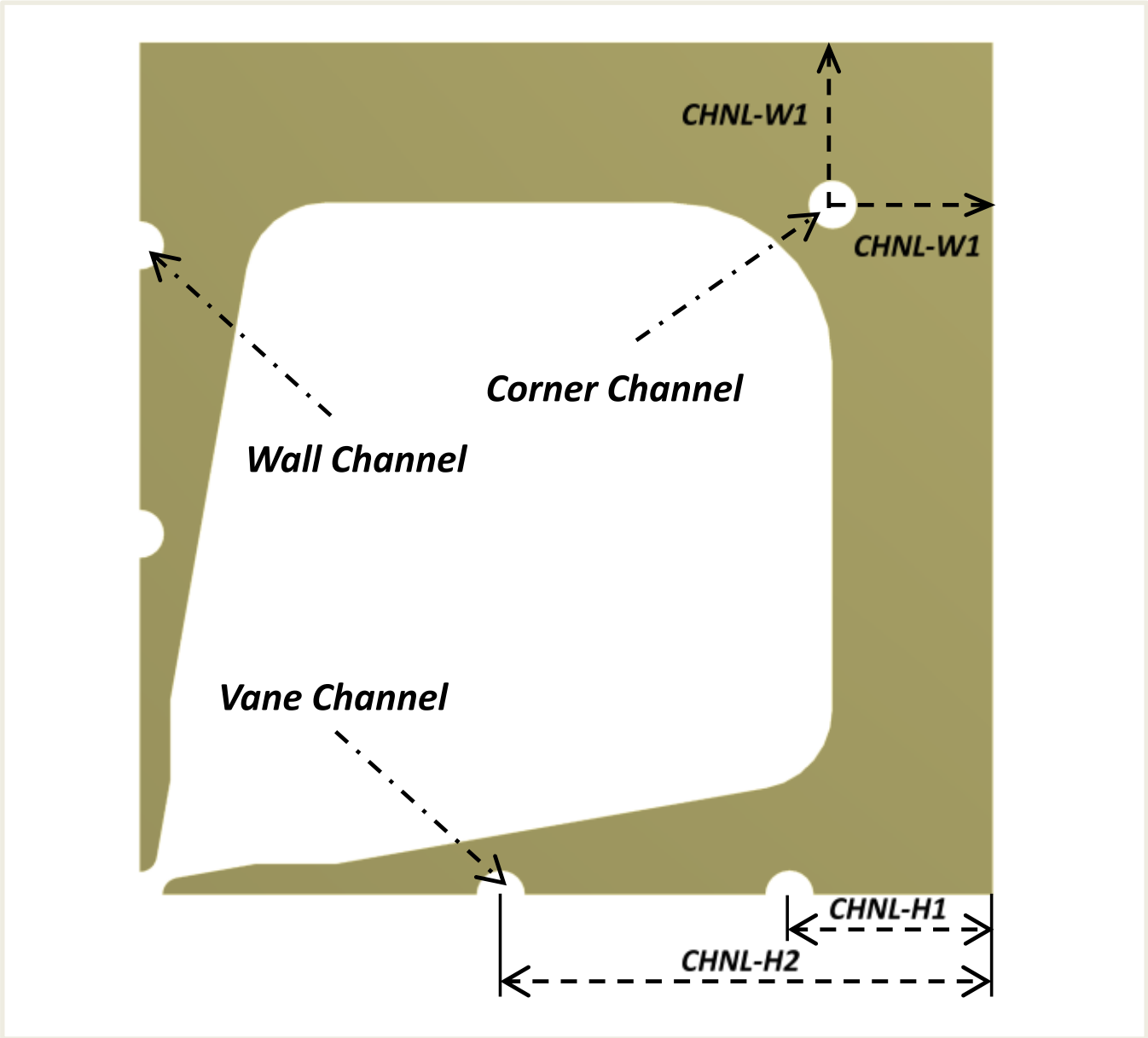

Fig.2 One-quarter of 10 mm RFQ slice model with openings of cooling channels for multiphysics simulations is adopted, due to structural symmetry; there are 2 vane channels close to the vane-tips, 2 wall channels put around the bottom of vanes, and 1 corner channel placed near the cavity corner; the reference positions of cooling channels are located at the outer edges of cavity wall; and the geometrical positions of vane, wall and corner channels are CHNL-H2, CHNL-H1 and CHNL-W1, respectively.

The slice RFQ model used for the optimization of cooling channel configuration is shown in Figure 2. The cavity wall is made of Oxygen-Free High Conductivity (OFHC) copper, which is preferable for its higher thermal and electrical conductivity, better ductility, easier brazing and so on. The gun-drilled cooling channels can be divided into 3 parts: Vane Channel (CHNL-H2), Wall Channel (CHNL-H1) and Corner Channel (CHNL-W1), whose diameters are 12 mm. The velocity of water flow passing through each channel is set to 2.29 m/s. The temperatures of inlet cooling water and ambient air are supposed to be 30 ℃ and 20 ℃, respectively. We substitute the thermophysical characteristics of cooling water into Eq. (1) to (4) and solve them for the heat convection coefficient of each channel. At the temperature of 30 ℃, the heat convection coefficient is calculated as 9988 W/(m$^2$·℃). When varying temperature from 25 ℃ to 35 ℃, the Reynolds number is between 30753 and 37941. Because the Reynolds number is always exceeding 4000, we could know that the cooling flow in each channel is in turbulent regime, which would ensure effective heat transfer from cavity wall to water flow. Additionally, the heat convection coefficient between outer cavity surface and ambient air is approximately 10 W/(m$^2$·°C), which has negligible effect on thermal studies. In Table 2, the properties of cooling water in different channels are summarized.

Table 2 Summary of cooling water properties obtained with analytical formulas.

| Parameters | Vane, wall and corner channels |
|---|---|
| Inlet cooling water temperature (°C) | 30 |
| Cooling channel diameter (mm) | 12 |
| Cooling water velocity (m/s) | 2.29 |
| Cooling water flow rate (l/s) | 0.26 |
| Number of channels | 4, 4, 4 |
| Reynolds number | 34288.359 |
| Prandtl number | 5.292 |
| Nusselt number | 190.238 |
| Heat convection coefficient (W/m$^2$/°C) | 9988 |

With regard to optimizing the configuration of cooling channels, we will take into account the following important aspects:

(1) Minimizing max temperature rise -> minimizing structure deformation;

(2) Reducing temperature variation along RFQ profile -> minimizing yield stress;

(3) Keeping high tuning sensitivity of vane channels and wall channels -> minimizing temperature difference of coolant for tuning;

### 3.1 Optimization of Vane Channel (CHNL-H2)

The multiphysics analysis is carried out with a fully parametric slice RFQ model. We run a parameter sweep of vane channel position (CHNL-H2) from 68 mm to 108 mm with a step of 10 mm while keeping fixed positions of wall channel (CHNL-H1) and corner channel (CHNL-W1). The temperature distribution and temperature gradient along the cavity inner surface (from one vane-tip to another) are monitored, which are plotted in Figure 3. Figure 3 (a) shows the maximum temperature decreases linearly as CHNL-H2 increases. The closer the vane channel to the vane-tip, the lower temperature rise will be at the vane-tip. Figure 3 (b) shows that CHNL-H2 has a main influence on the temperature distribution on the RFQ vane, while the temperature distribution near cavity corner is almost independent of it. Figure 3 (c) shows that the variation of temperature gradient along cavity profile could be smaller, when the vane channels are closer to the vane-tips. Therefore CHNL-H2 is set to 108 mm, which will leave some safe margin for the opening Pi-mode field stabilizer (PISL) hole in the RFQ vane.

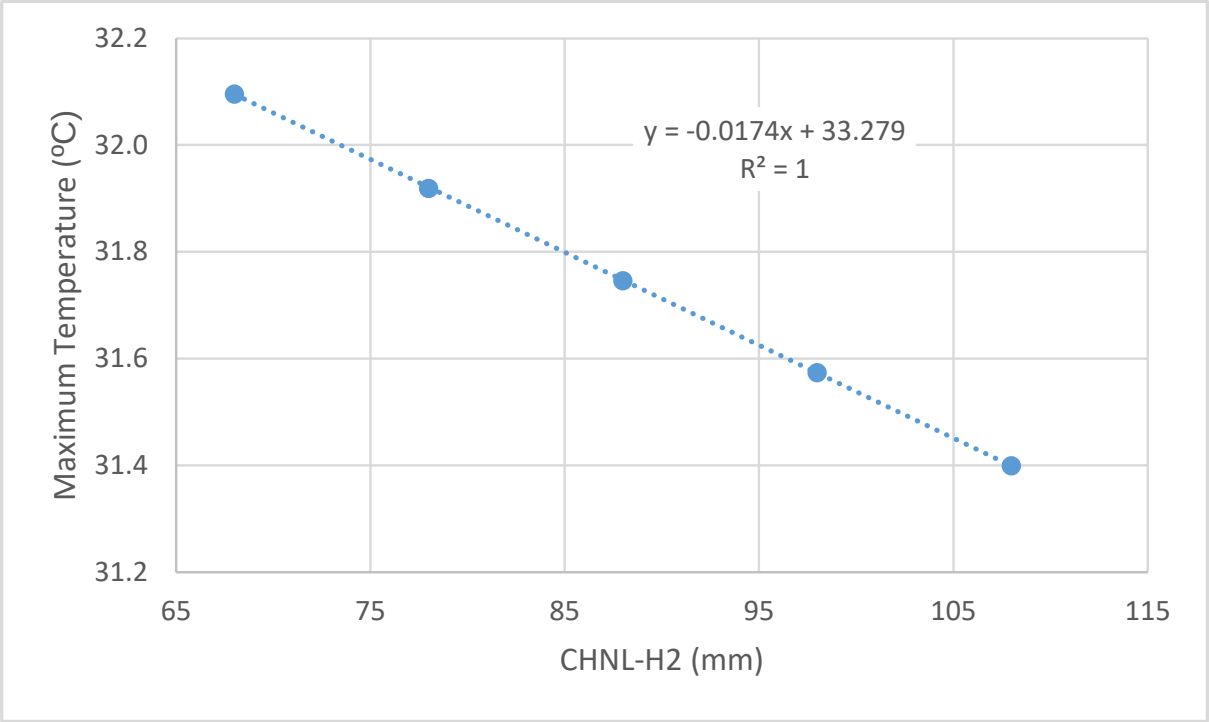

(a)

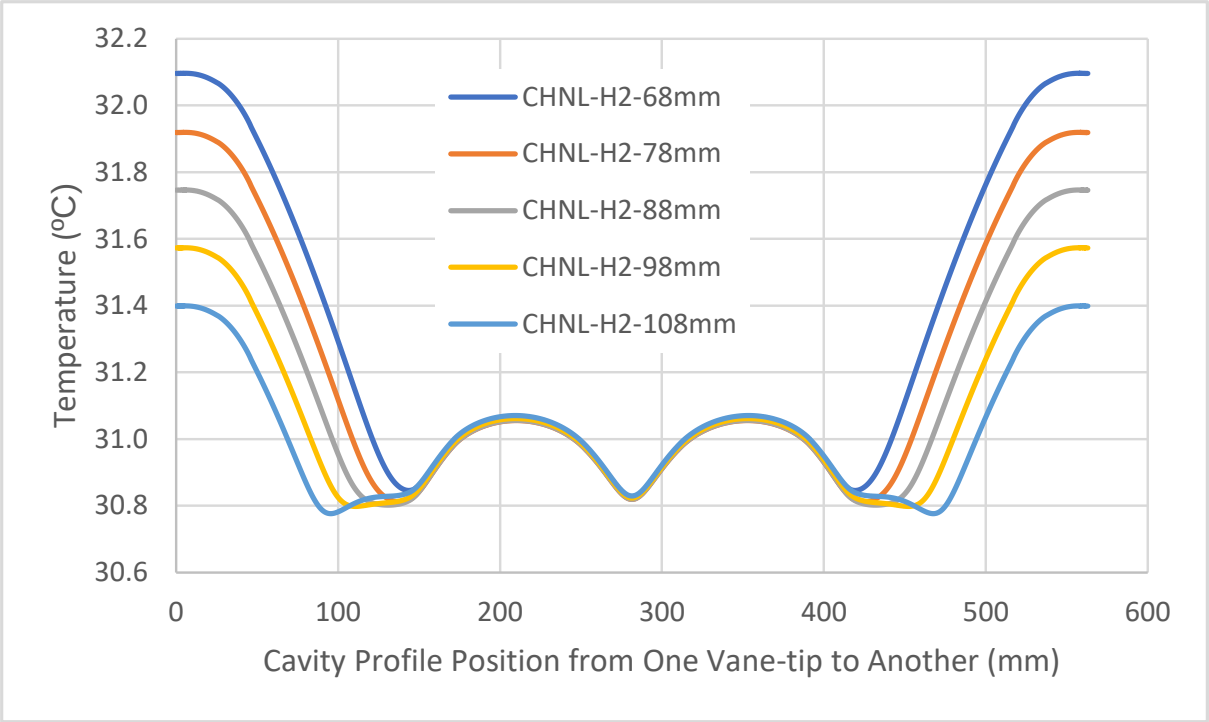

(b)

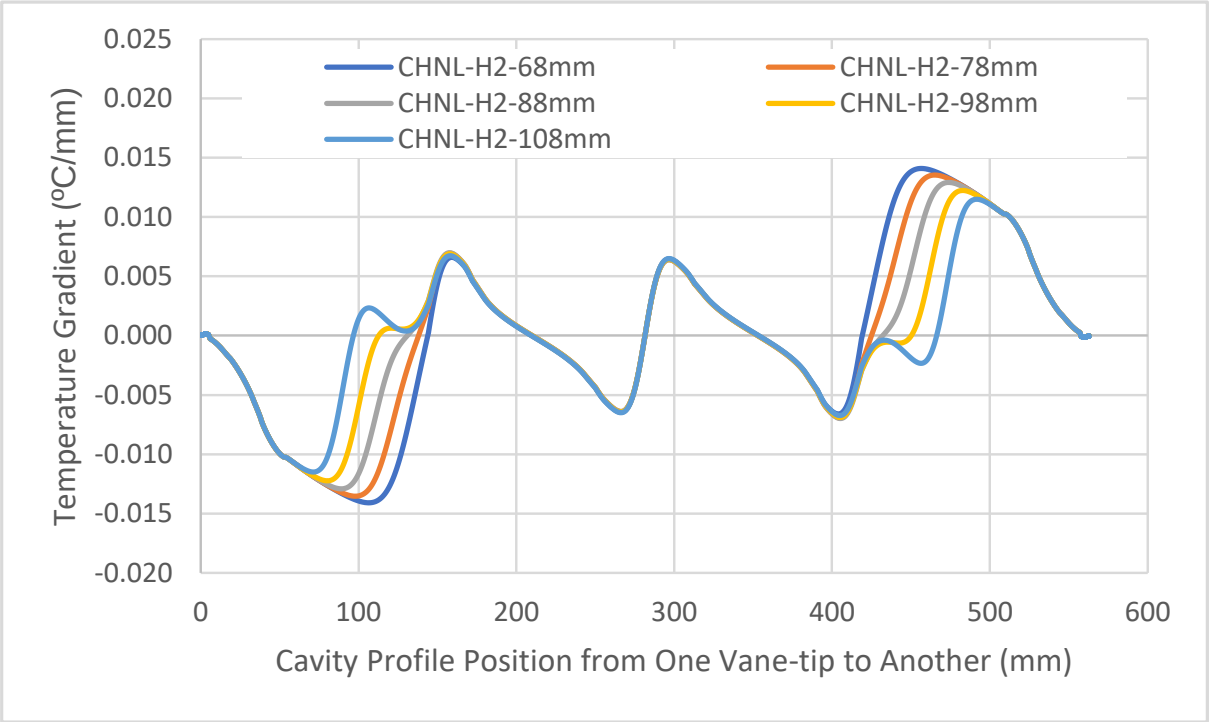

(c)

Fig.3 Evolutions of (a) cross-section's maximum temperature, (b) temperature distribution, (c) temperature gradient (ºC/mm) along

cavity inner profile with different vane channel positions (CHNL-H2) from 68 to 108 mm with a step of 10 mm.

### 3.2 Optimization of Wall Channel (CHNL-H1)

Next, we optimize the position of wall channel (CHNL-H1) with a range of 30 to 70 mm in a similar way. The simulated results are shown in Figure 4. Figure 4 (a) shows that the maximum temperature is located at the vane-tip and is almost linear with the wall channel position. Figure 4 (b) shows that the variation of temperature distribution from one vane corner to another is dominated by CHNL-H1. But Figure 4 (c) suggests the variation of temperature gradient is localized near vane corner. So, the wall channel position (CHNL-H1) is determined to be 50 mm for a relatively smaller maximum temperature rise and variation of temperature gradient.

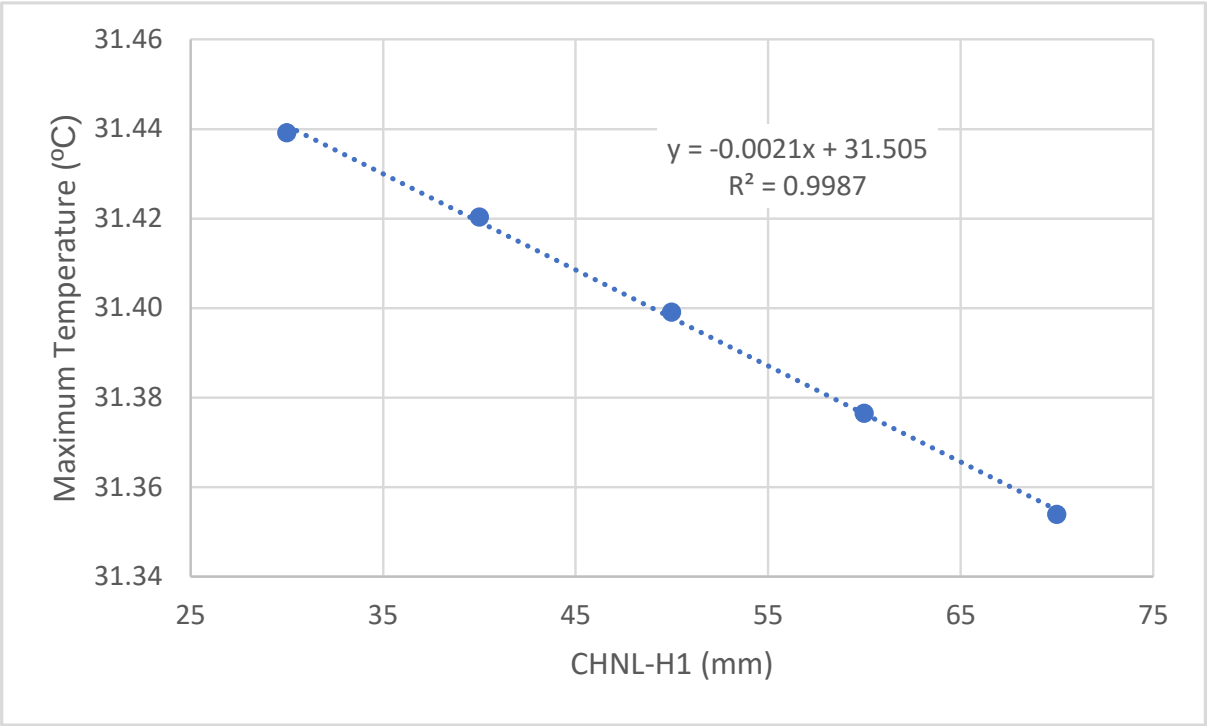

(a)

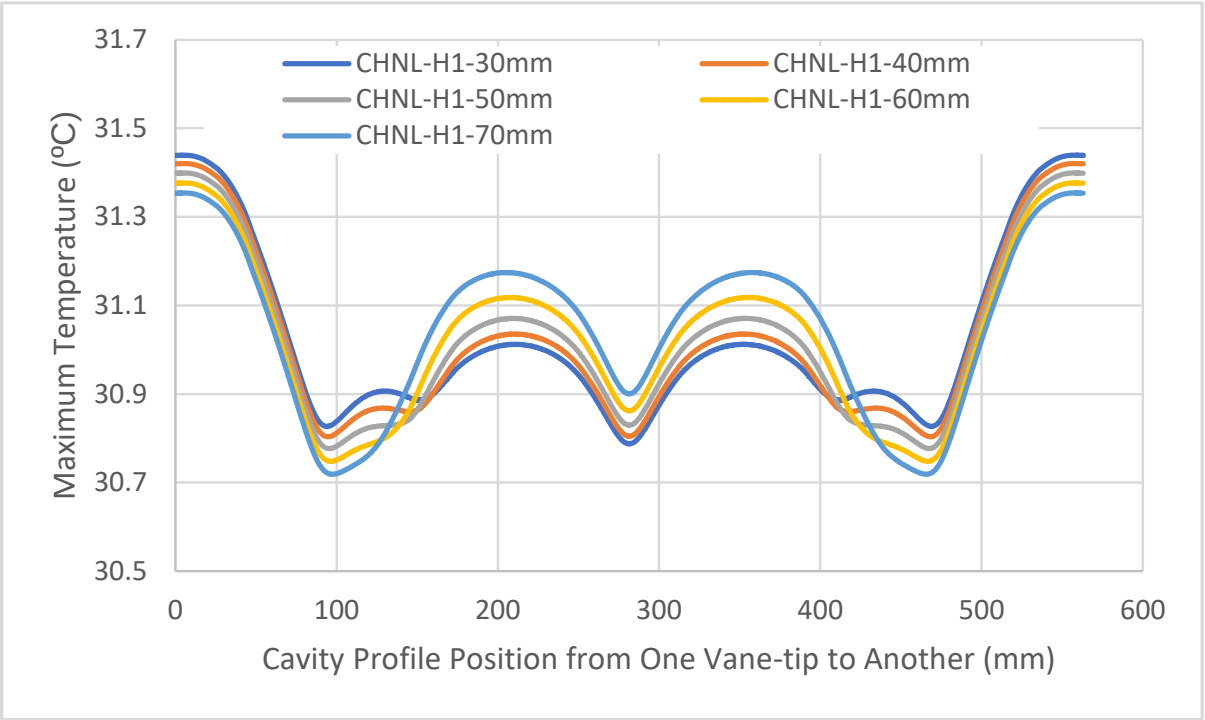

(b)

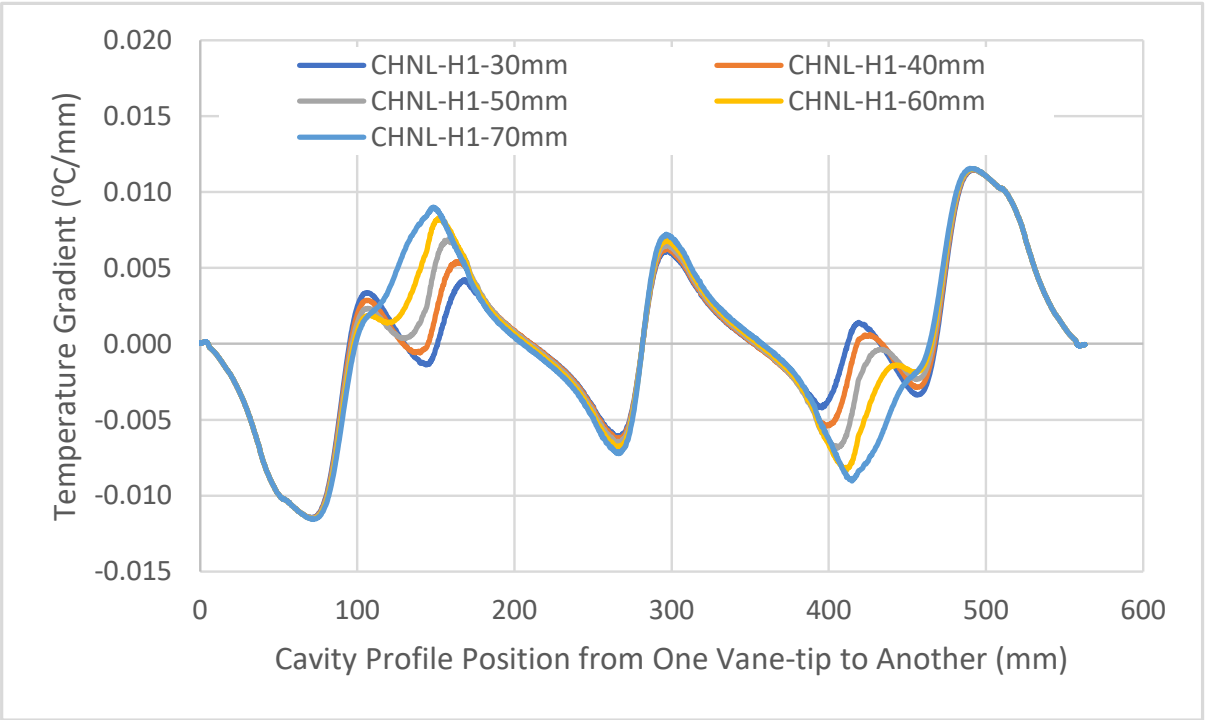

(c)

Fig. 4 Evolutions of (a) cross-section's maximum temperature, (b) temperature distribution, (c) temperature gradient (ºC/mm) along cavity inner profile with different wall channel positions (CHNL-H1) ranging from 30 to 70 mm with a step of 10 mm.

### 3.3 Optimization of Corner Channel (CHNL-W1)

Then, a series of simulations are performed for the optimization of corner channel position (CHNL-W1) from 35 to 43 mm. As shown in Figure 5 (a), the maximum temperature located at the vane-tip is insensitive to the corner channel position (CHNL-W1). Figure 5 (b) and (c) demonstrate that the changes of both temperature distribution and temperature gradient are local effects confined around cavity corner. Although the temperature variation is relatively small in the case of CHNL-W1 = 35 mm, the choice of 39 mm will give more space for cavity brazing between the major vane and minor vane.

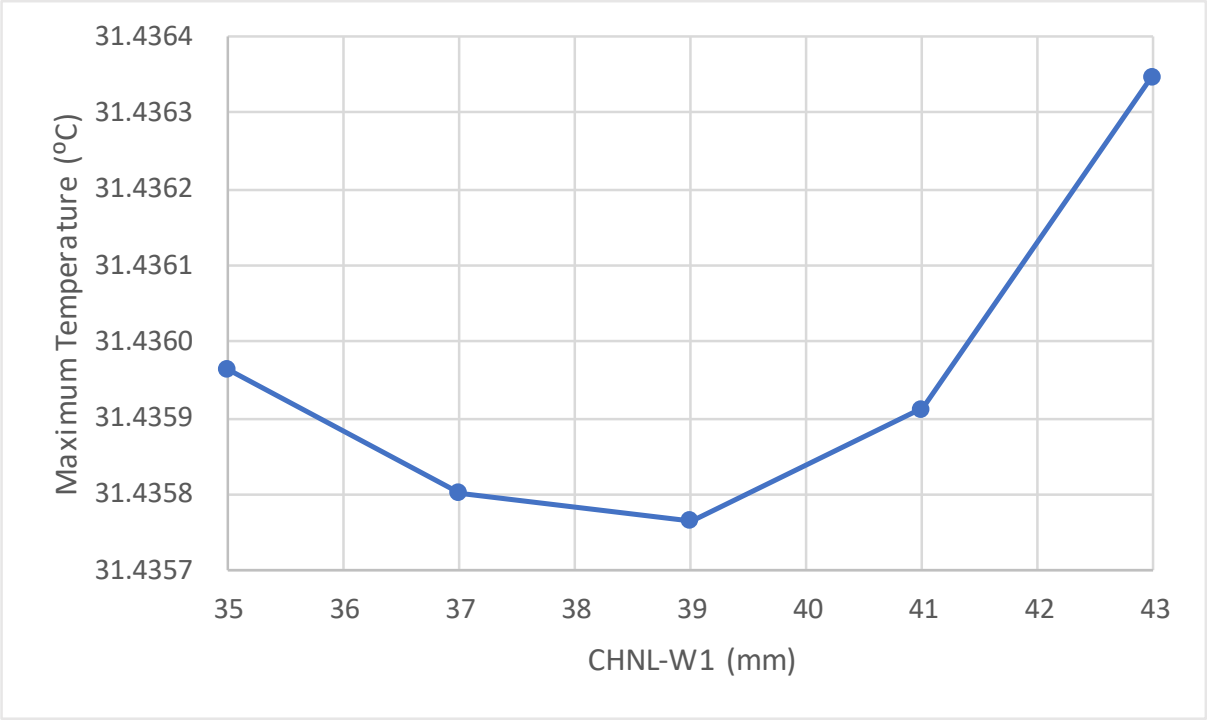

(a)

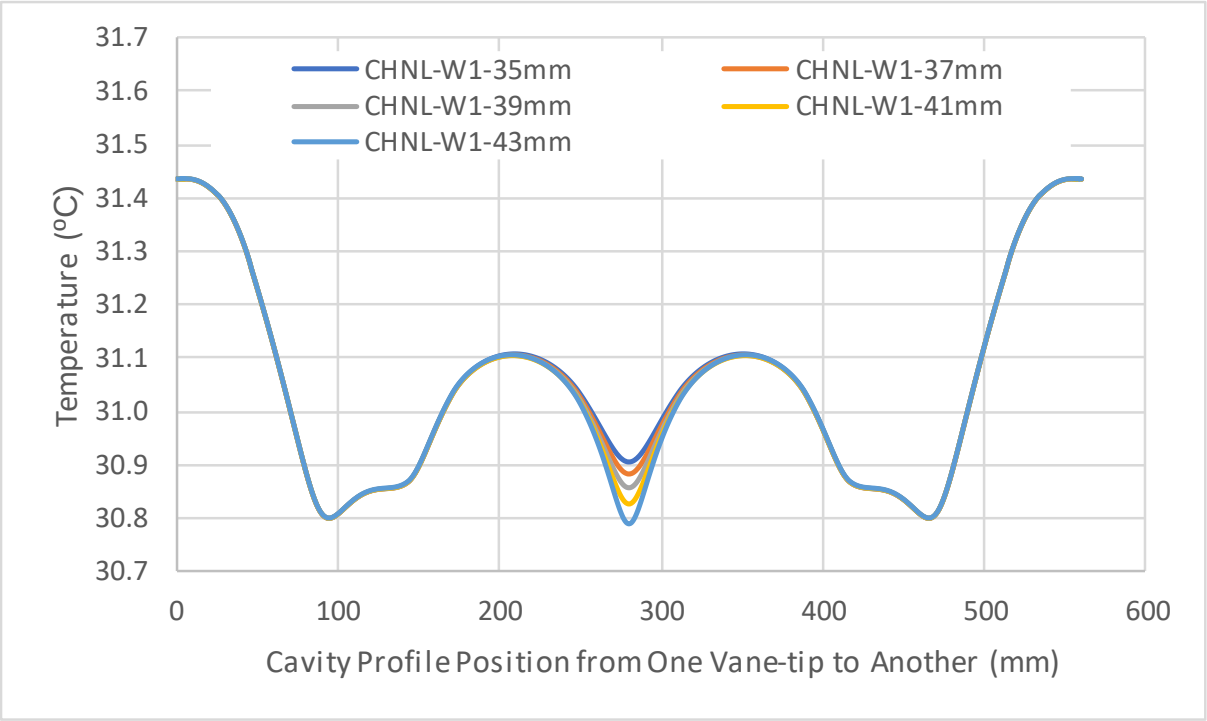

(b)

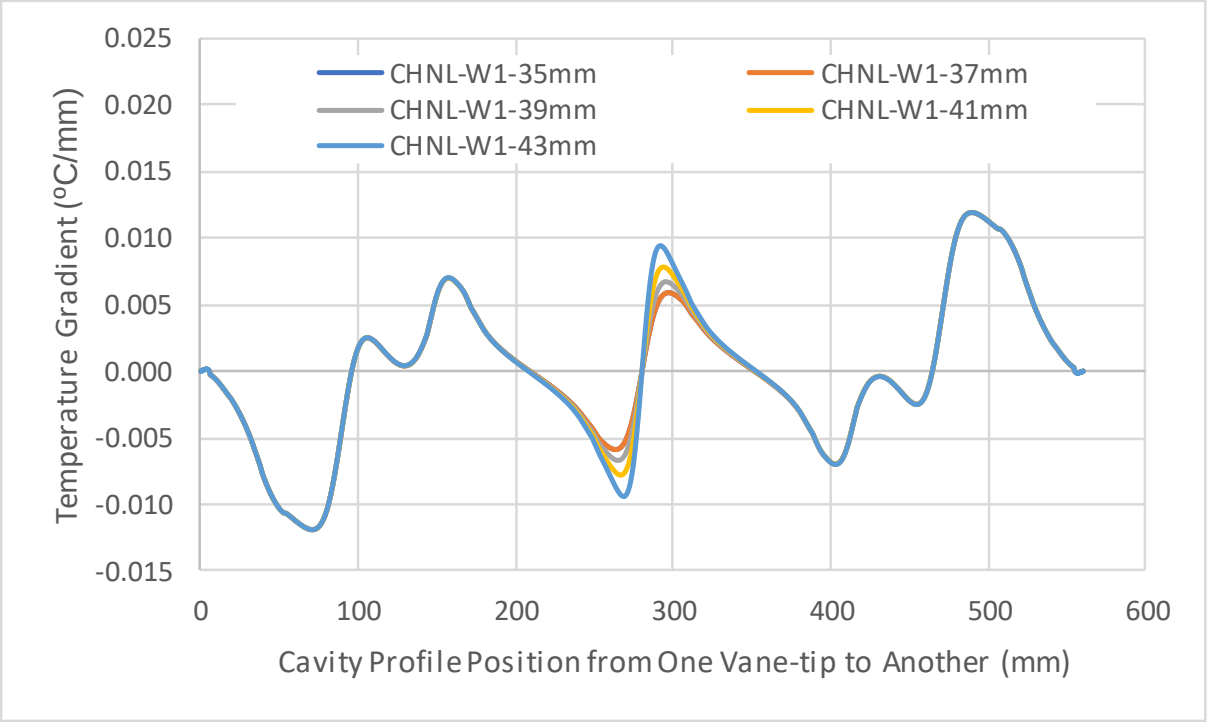

(c)

Fig.5 Evolutions of(a) cross-section's maximum temperature, (b) temperature distribution, (c) temperature gradient (°C/mm) along cavity inner profile with different corner channel positions (CHNL-W1) in the range from 35 to 43 mm with an increment of 2 mm.

### 3.4 Tuning Sensitivity of Cooling Channels

The preliminary optimizations of cooling channel configurations having been done, in this part, the frequency detuning results from RF heating predicted via multiphysics analysis will be presented. To simplify the design of cooling system and the control of cooling water temperature, the vane channel and wall channel will be grouped as one. Different frequency responses can be obtained by adjusting the temperature of cooling water passing in vanes and walls. When calculating the tuning sensitivity coefficient of vane channels, we only vary the temperature ($T_{vane}$) of cooling water in vanes, while cooling water temperature in walls ($T_{wall}$) is fixed at 30 ℃, and vice versa. The tuning sensitivity of different configurations of cooling channel positions are summarized in Table 3.

Table 3 Summary of tuning sensitivity of vane, wall and corner channels in steady-state analysis with 10% duty cycle.

| Case: Optimization of CHNL-H1 | | | | | |
|---|---|---|---|---|---|
| H1 (mm) | H2 (mm) | W1 (mm) | $\partial f/\partial T_{vane}$ (kHz/°C) | $\partial f/\partial T_{wall}$ (kHz/°C) | Difference (kHz/°C) |
| 70 | 108 | 39 | -15.1 | 10.9 | -4.2 |
| 60 | 108 | 39 | -16.0 | 11.8 | -4.2 |
| 50 | 108 | 39 | -16.8 | 12.7 | -4.1 |
| 40 | 108 | 39 | -17.6 | 13.4 | -4.2 |
| 30 | 108 | 39 | -18.2 | 14.1 | -4.1 |
| Case: Optimization of CHNL-H2 | | | | | |
| H1 (mm) | H2 (mm) | W1 (mm) | $\partial f/\partial T_{vane}$ (kHz/°C) | $\partial f/\partial T_{wall}$ (kHz/°C) | Difference (kHz/°C) |
| 50 | 108 | 39 | -16.8 | 12.7 | -4.1 |
| 50 | 98 | 39 | -16.2 | 12.0 | -4.2 |
| 50 | 88 | 39 | -15.3 | 11.2 | -4.1 |
| 50 | 78 | 39 | -14.1 | 10.0 | -4.1 |
| 50 | 68 | 39 | -12.7 | 8.5 | -4.2 |
| Case: Optimization of CHNL-W1 | | | | | |
| H1 (mm) | H2 (mm) | W1 (mm) | $\partial f/\partial T_{vane}$ (kHz/°C) | $\partial f/\partial T_{wall}$ (kHz/°C) | Difference (kHz/°C) |
| 50 | 108 | 43 | -16.9 | 12.7 | -4.2 |
| 50 | 108 | 41 | -16.9 | 12.7 | -4.2 |
| 50 | 108 | 39 | -16.8 | 12.7 | -4.1 |
| 50 | 108 | 37 | -16.8 | 12.6 | -4.2 |
| 50 | 108 | 35 | -16.8 | 12.6 | -4.2 |

As can be seen from the table above, we can conclude that:

(1) Tuning coefficients of '$\partial f/\partial T_{vane}$' and '$\partial f/\partial T_{wall}$' have opposite signs -> will give opposite frequency response;

(2) Difference between '$\partial f/\partial T_{vane}$' and '$\partial f/\partial T_{wall}$' is fixed -> are independent of channel layouts;

(3) Increasing CHNL-H1, $|\partial f/\partial T_{vane}|$ sensitivity reduces (↓), and so does $|\partial f/\partial T_{wall}|$ sensitivity (↓);

(4) Raising CHNL-H2, $|\partial f/\partial T_{vane}|$ sensitivity increases (↑), and so does $|\partial f/\partial T_{wall}|$ sensitivity (↑);

(5) Tuning sensitivity of '$\partial f/\partial T_{vane}$' and '$\partial f/\partial T_{wall}$' are independent of 'CHNL-W1' position;

The channel configuration (CHNL-H1, CHNL-H2, CHNL-W1) determined in Section 3.1 to 3.4 to be a combination of (50 mm, 108 mm, 39 mm), thus not only gives a cooling design with high tuning sensitivity in vanes and walls to afford dynamic tuning, but also decreases the temperature rise and the variation of temperature gradient.

The temperature distribution computed with inlet cooling water temperatures of 30 ℃ in steady-state thermal analysis is shown in Figure 6. The maximum temperature is found to be at the vane-tip, reaching 31.41 ℃.

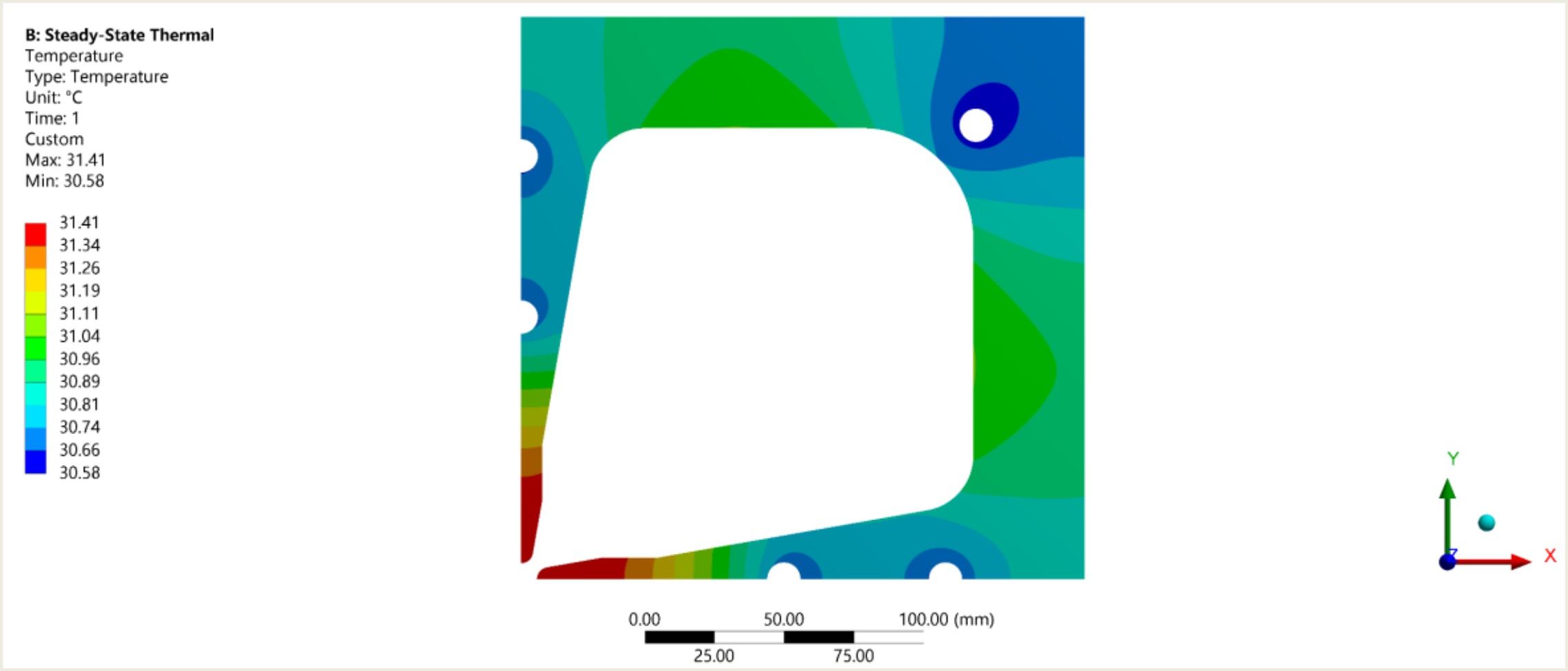

Fig.6 Steady-state temperature map (in °C) simulated with ANSYS code at a duty factor of 10%.

The temperature distribution calculated in the thermal study is transferred to the structural analysis. Then, the attained thermal displacement pattern is shown in Figure 7. The maximum deformation is evaluated to be 5.5 μm, which is located at the cavity corner. The RFQ von-Mises stress map is displayed in Figure 8, and its maximum stress is calculated as 2.886 MPa, whose stress level is well within the safe limit of copper.

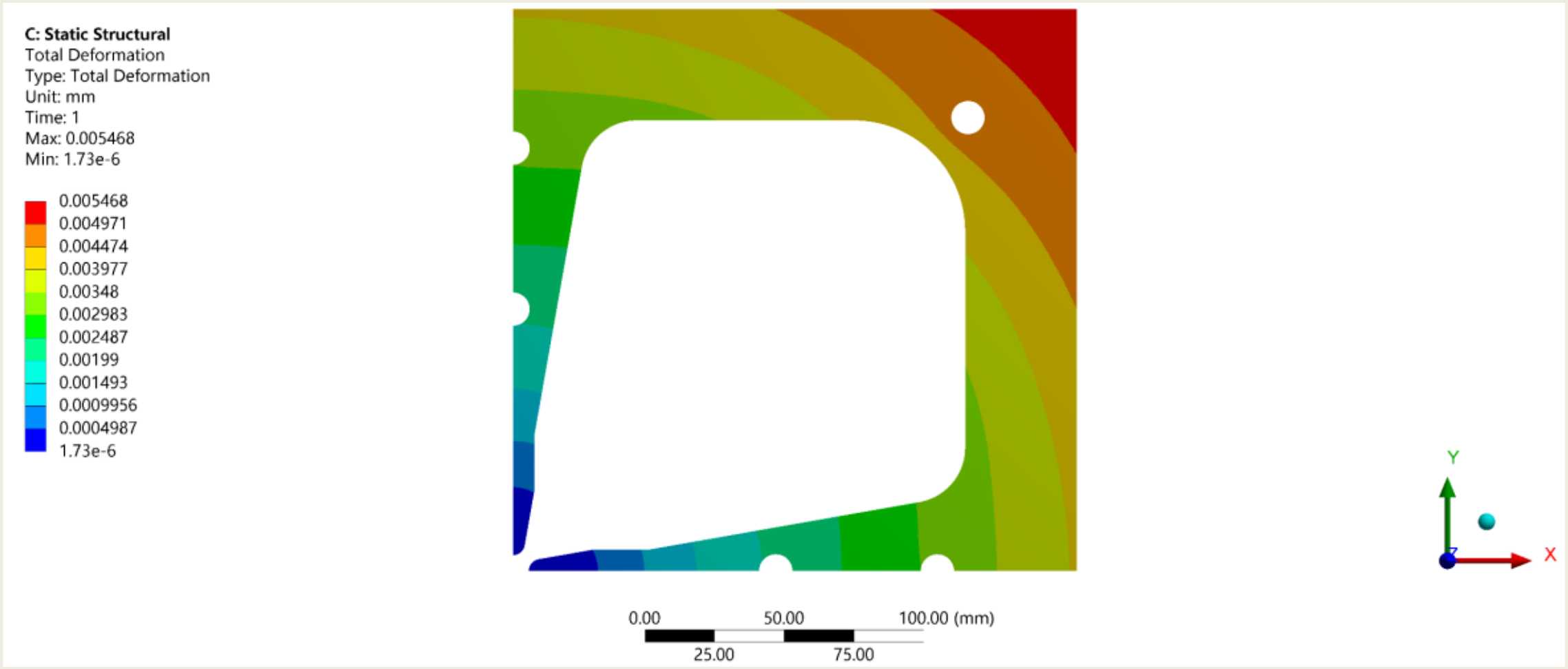

Fig.7 Resulting cavity deformation (in mm) induced by thermal load.

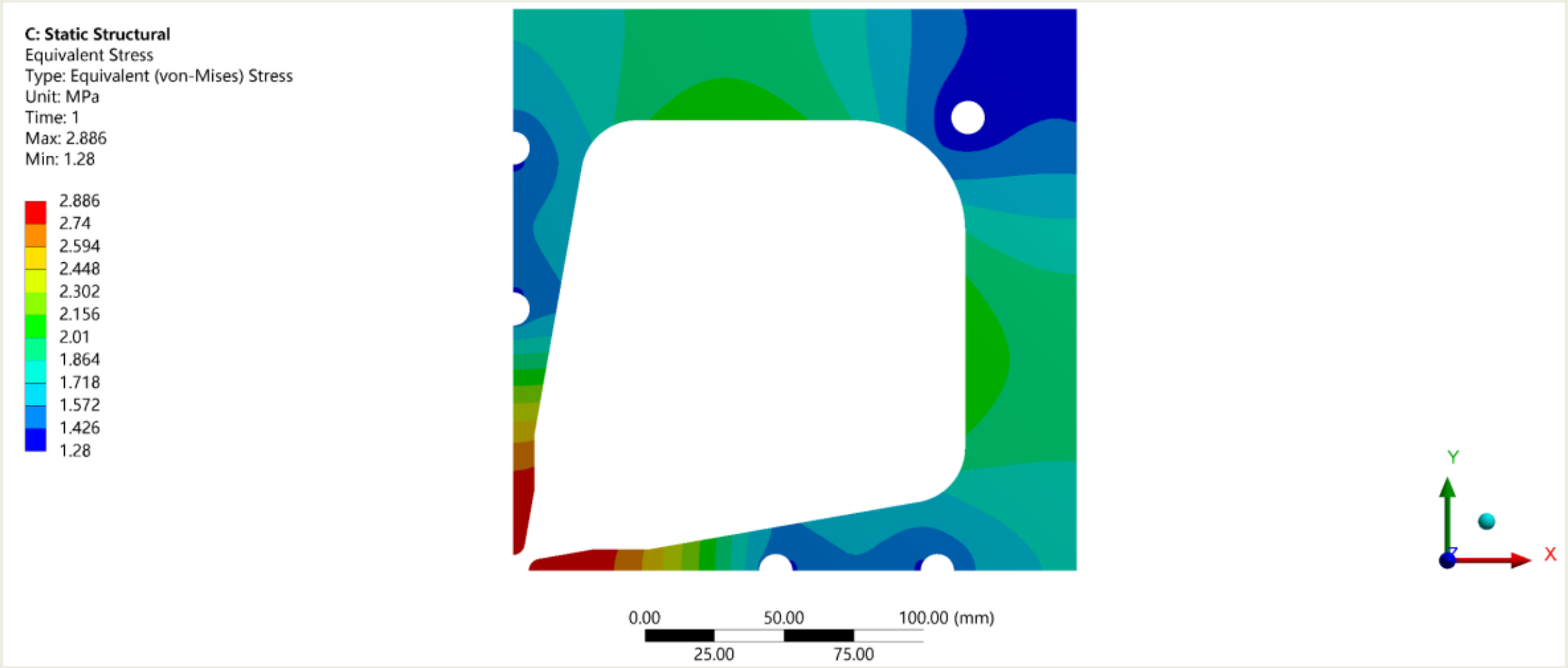

Fig.8 Yield stress distribution (in MPa) of slice RFQ.

The frequency responses due to the adjustments of inlet $T_{vane}$ from 24 to 36 °C is illustrated in Figure 9. After linear fitting, the slope '$\partial f/\partial T_{vane}$' of the curve is obtained, and in the same way we could solve for '$\partial f/\partial T_{wall}$' in the walls. They are calculated as -16.8 kHz/°C and 12.7 kHz/°C, respectively, which will be used as inputs to the resonance control cooling system (RCCS) for frequency correction during high power operation.

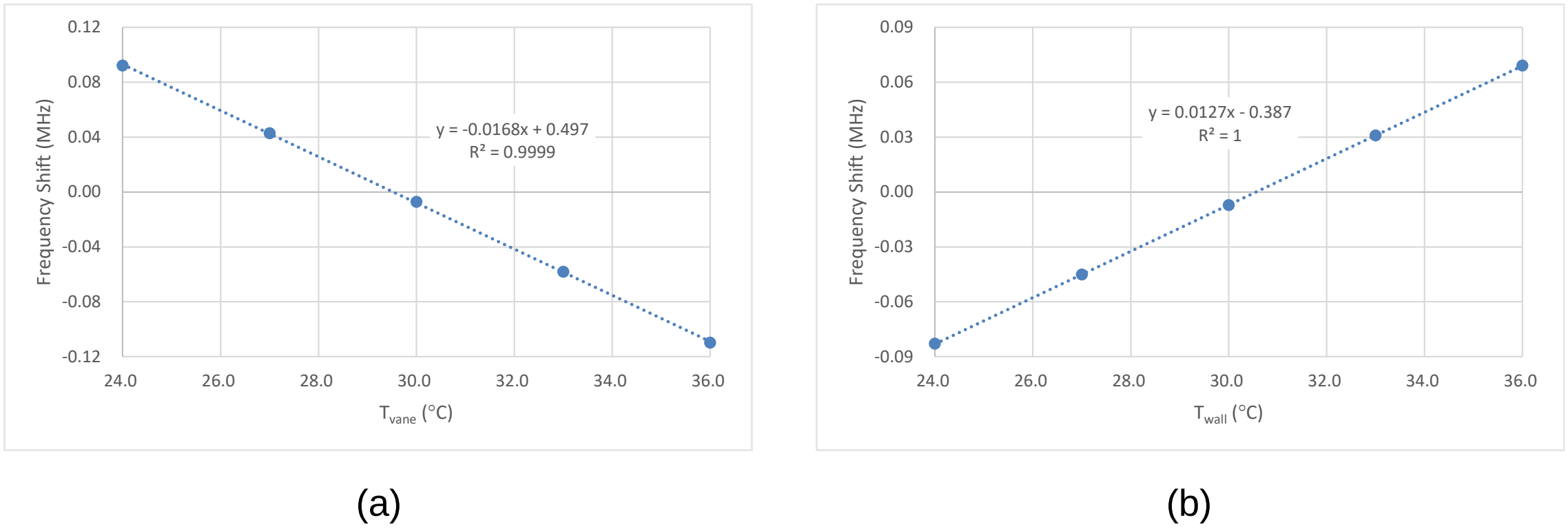

Fig.9 Resonant frequency responses when applied with different inlet cooling water temperatures (a) $T_{vane}$ and (b) $T_{wall}$, whose slopes of curves indicate their corresponding tuning sensitivity in vanes and walls.

A benchmark study carried out using CST and ANSYS codes is summarized in Table 4. Both codes give consistent results. Slight differences might be explained by different techniques for mesh generations.

Table 4 Benchmark results obtained from CST and ANSYS code packages.

| Parameter | CST | ANSYS |
|---|---|---|
| Ambient temperature (°C) | 20 | 20 |
| Cooling water temperature (°C) | 30 | 30 |
| Cooling channel diameter (mm) | 12 | 12 |
| Velocity of water flow (m/s) | 2.29 | 2.29 |
| Heat convection coefficient (W/m$^2$ /°C) | 9988 | 9988 |
| Power Scaling Factor (Joule) | 0.000950 | 0.000950 |
| Maximum surface loss density (W/m$^2$) | 1707.0 | 1714.2 |
| Maximum temperature rise (°C) | 1.39 | 1.41 |
| Minimum temperature rise (°C) | 0.59 | 0.58 |
| Maximum deformation (μm) | 5.505 | 5.468 |
| Maximum yield stress (MPa) | 2.837 | 2.886 |
| Minimum yield stress (MPa) | 1.301 | 1.280 |
| $\partial f/\partial T_{vane}$ (kHz/°C) | -16.8 | -16.8 |
| $\partial f/\partial T_{wall}$ (kHz/°C) | 12.7 | 12.7 |

## 4. Transient Analysis during Start-up

### 4.1 Time Constant and Thermal Equilibrium

As mentioned earlier, the inlet cooling water temperature and background temperature are assumed to be 30 °C and 20 °C, respectively. With the purpose of reducing frequency detuning during commissioning, the RF power can be fed when a thermal equilibrium state is reached between cooling water and cavity wall.

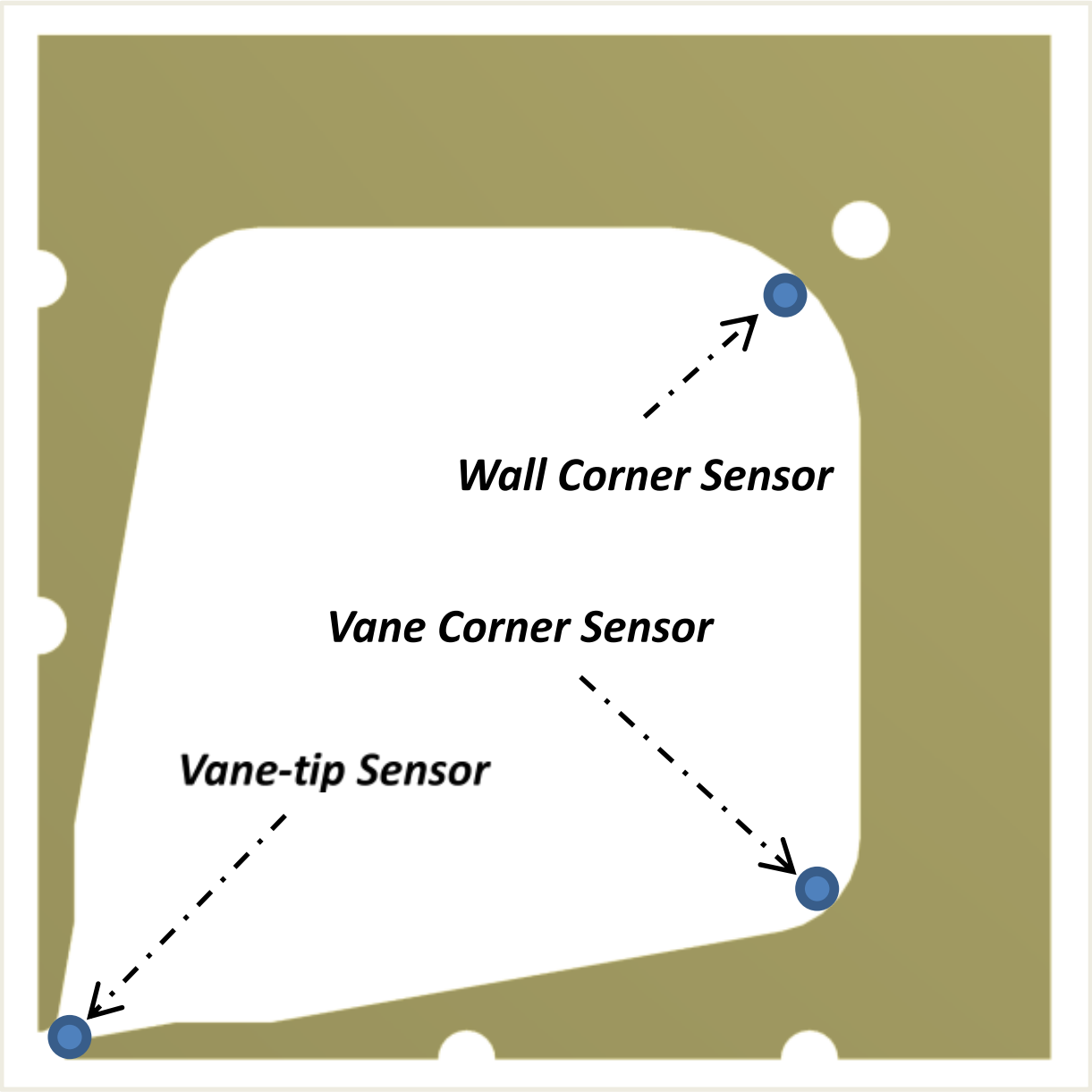

Fig.10 FEM model with temperature sensors placed at the vane-tip, vane corner, and wall corner.

Figure 10 shows FEM model used for studying transient response, and three temperature sensors separately located at the vane-tip, vane corner and wall corner are used to track the temperature variation behavior before thermal equilibrium (see Figure 11). Because the temperature evolutions at the sensors show exponential-like growth processes, we can take advantage of the definition of rising time constant in electronics,

$$t_r = t_{90\%} - t_{10\%} = \tau \cdot ln(9) \quad (5)$$

Then time constants at these three locations can be given as 17.3 s ($\tau_{vane\text{-}tip}$), 27.3 s ($\tau_{vane\text{-}corner}$), and 12.3 s ($\tau_{wall\text{-}corner}$), respectively. And we can know that after about 150 seconds (~ 5 times $\tau_{vane\text{-}corner}$), the RFQ cavity will reach thermal equilibrium with cooling water.

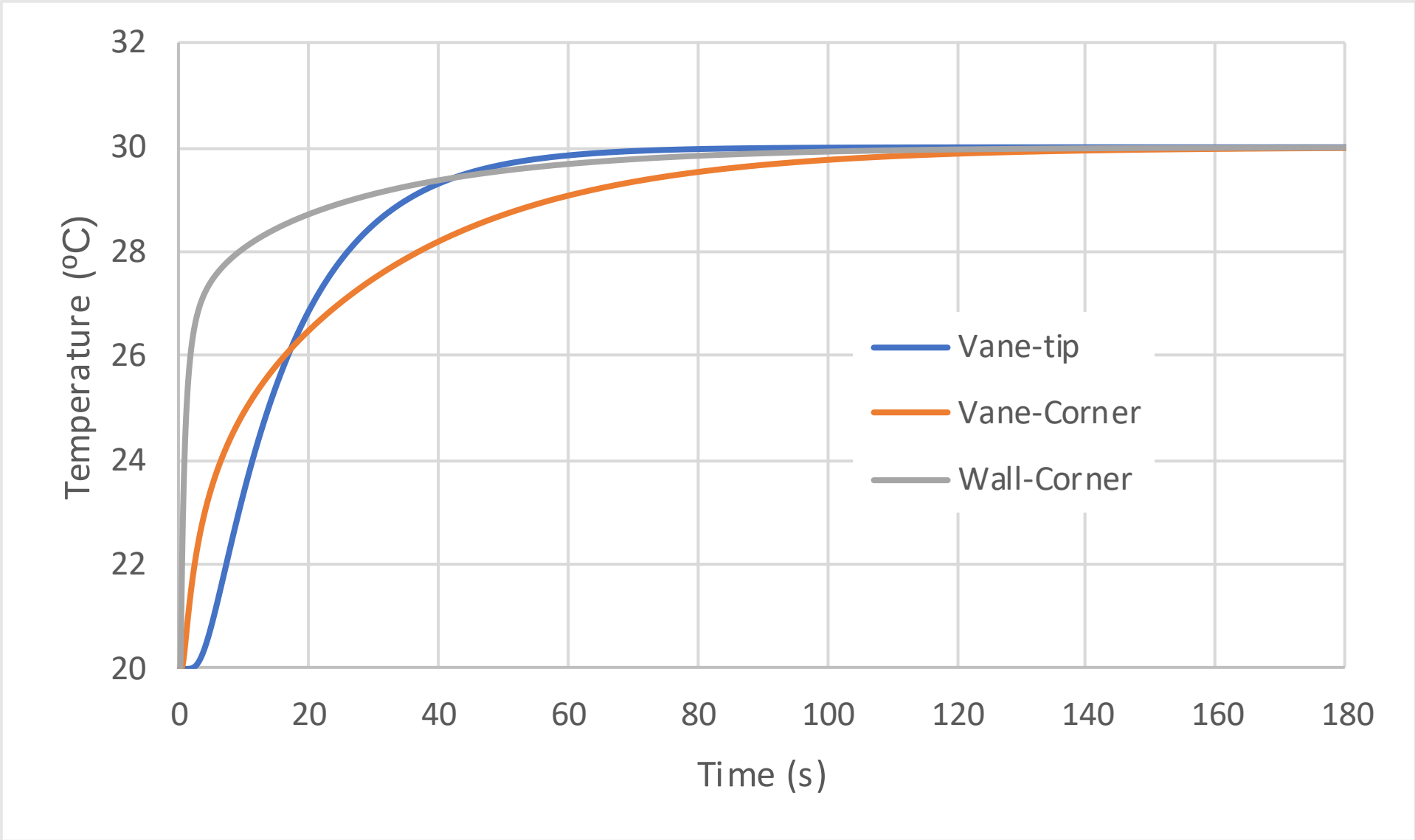

Fig.11 Transient temperature variations recorded by sensors at the vane-tip, vane corner and wall corner as a function of time; all sensor temperatures start from ambient temperature of 20 °C, and finally equal to the temperature of cooling water at 30 °C.

### 4.2 RF-on Transient

In next simulations, a series of transient analyses have been performed utilizing an initial temperature distribution of thermal equilibrium obtained in Section 4.1. And the same FEM model is carried out in simulations to characterize the frequency response during RF power ramp-up. However, the input growth rate of RF power should be constrained because the resonance control cooling system is not able to realize the fast frequency correction. The according frequency response is plotted in Figure 12. As it shows, the frequency drop is due to the RF heating on the cavity walls and it roughly takes 750 seconds to achieve a state of thermal equilibrium with RF power on. After that, slightly adjusting the temperature of cooling water in the vane channels will tune the resonant frequency to the designed one. Considering a cold start of the RFQ accelerator, it will take approximate 15 minutes (or 900 seconds) to get into a stable operation.

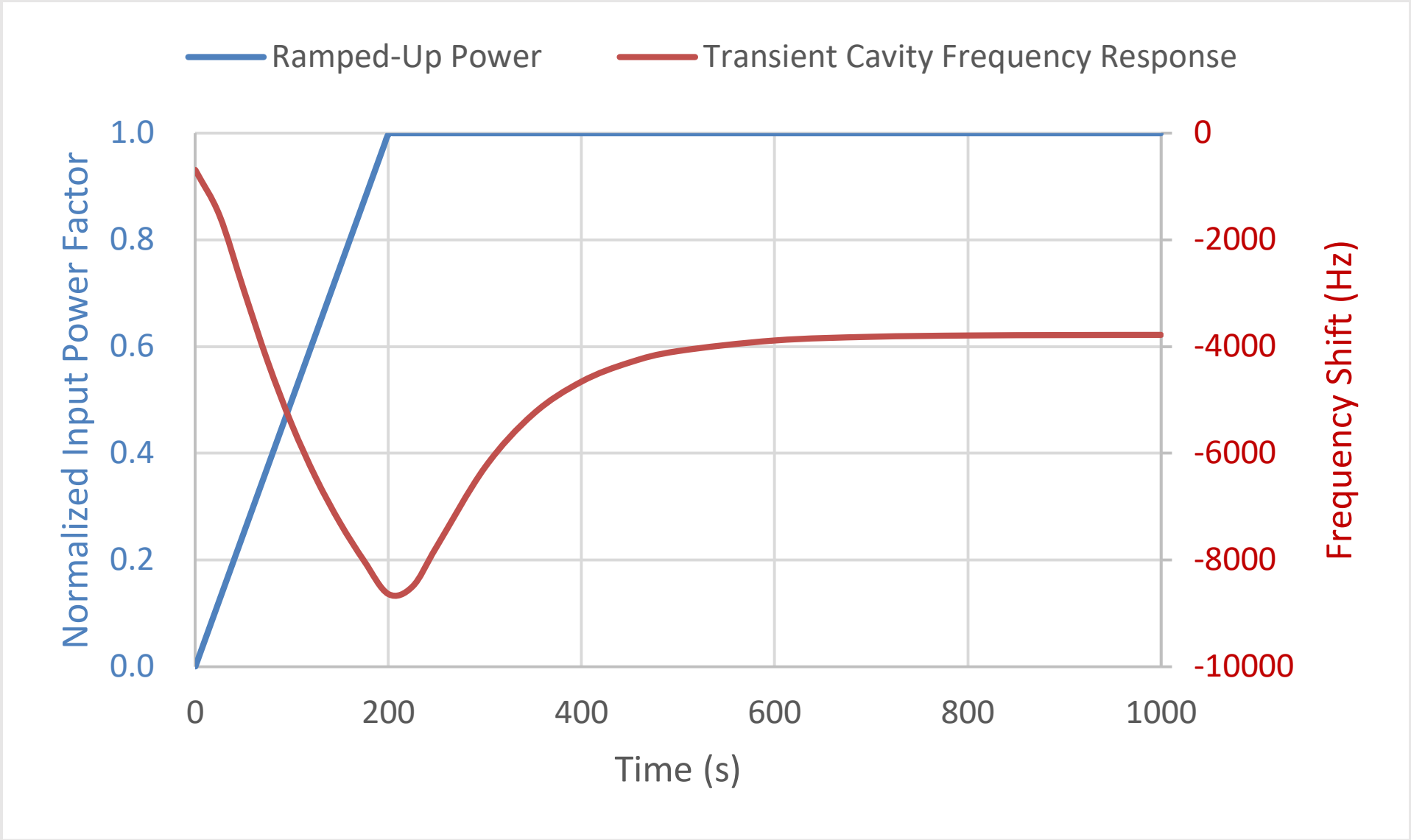

Fig.12 Evolutions of (left) Normalized input power factor and (right) transient frequency response induced by thermal load.

## 5. Conclusion

We have proposed a method to optimize the cooling channel positions by minimizing max temperature rise, reducing temperature gradient variation, and keeping a relatively high tuning sensitivity of vane and wall channels. A series of steady-state multiphysics analyses have been studied to predict temperature map, structure deformation as well as frequency detuning at nominal operation condition. In the meantime, a course of transient simulations show how frequency response evolves during power-on from cold start and how much time is needed for a cavity to reach thermal equilibrium. By alternating temperature ($T_{vane}$) in the vane channels while fixing $T_{wall}$ in the wall channels, the cavity resonant frequency can be adjusted to the designed one.